# Pathogen Evolution During Vaccination Campaigns

Troy Day[1], David A. Kennedy[2], Andrew F. Read[2,3], Sylvain Gandon[4]

1. Department of Mathematics and Statistics, Department of Biology, Queen's University, Kingston, K7L 3N6, Canada. day@queensu.ca
2. Department of Biology, Pennsylvania State University, University Park, PA 16802, USA
3. Department of Entomology, Pennsylvania State University, University Park, PA 16802, USA
4. CEFE, CNRS, Univ Montpellier, EPHE, IRD, Montpellier, France

**With the unprecedented global vaccination campaign against SARS-CoV-2 attention has now turned to the potential impact of this large-scale intervention on the evolution of the virus. In this perspective we summarize what is currently known about evolution in the context of vaccination from research on other pathogen species, with an eye towards the future evolution of SARS-CoV-2.**

It is useful to think of the temporal dynamics of evolutionary change for novel pathogens like SARS-CoV-2 as passing through two phases. In the first phase the host population is immunologically naïve and selection strongly favours adaptation to these abundant naïve hosts. In the second phase a growing proportion of the population will have an immunological history with the pathogen, either through natural infection or vaccination, and thus selection will shift, increasingly favouring adaptation to these hosts. In this article we will focus primarily on pathogen evolution in response to vaccination but return to the issue of evolution driven by immunity acquired from natural infections in our conclusions.

Pathogen adaptation to naïve and vaccinated hosts depends on the appearance of new variants as well as on their fitness in each host type. We can quantify fitness by considering both the absolute per capita growth rate of infections caused by a variant as well as this growth rate relative to the growth rate of the currently dominant type (sometimes called the wildtype). The absolute growth rate will determine if the variant can spread in a population while the relative growth rate will determine if the variant can increase in frequency and thereby potentially displace the currently dominant type.

For a variant to spread in a population its absolute growth rate must be positive (equivalently, its reproduction number must be larger than one). The absolute growth rate, $r_i$, of infections caused by any pathogen variant $i$ can be approximated as (**Appendix 1**)

$$r_i = (1 - p)r_{i,N} + pr_{i,V} \tag{1}$$

where $p$ is the fraction of the population vaccinated, and $r_{i,N}$ and $r_{i,V}$ are the growth rates of infections by variant $i$ in a fully naïve and fully vaccinated population, respectively [1, 2].

For a variant to increase in relative *frequency*, and thus potentially displace the wildtype, its selection coefficient, *s*, defined as the difference between its growth rate and that of the wildtype, must be positive. For the above model this selection coefficient is given by

$$s = (1-p)\Delta r_N + p\Delta r_V \qquad (2)$$

where $\Delta r_N$ and $\Delta r_V$ are the differences in growth rate between the variant and the wildtype in a fully naïve and fully vaccinated population, respectively.

With this we can give a precise definition of a variant being adapted to vaccinated or naïve host populations. If $\Delta r_V > 0$ then the variant is more fit (i.e., has a higher growth rate) than the wildtype in a population of vaccinated hosts and so we say it is adapted to vaccinated host populations (equivalently, it is vaccine-adapted). Likewise, if $\Delta r_N > 0$ then the variant is more fit (i.e., has a higher growth rate) than the wildtype in a population of naïve hosts and so we say it is adapted to naïve host populations. Thus, in the first phase of an outbreak, when the fraction of vaccinated hosts $p$ is small, selection strongly favours variants for which $\Delta r_N > 0$ while, in the second phase, when $p$ is large, it strongly favours variants for which $\Delta r_V > 0$. In what follows we will focus on vaccine-adapted variants (i.e., those for which $\Delta r_V > 0$). Note that while there are many molecular and cellular mechanisms playing out within an infected host that can make a variant vaccine-adapted (**Box 1),** it is the impact of these mechanisms on the growth rate of the population of infected hosts that determines whether a variant spreads.

The above ideas lead to two useful ways of categorizing vaccine-adapted variants. First, if a vaccine-adapted variant is also adapted to naïve host populations (i.e., $\Delta r_N > 0$) then we refer to it as a "generalist" variant since it is better at spreading than the wildtype irrespective of host type. Conversely, if a vaccine-adapted variant is maladapted to naïve host populations (i.e., $\Delta r_N < 0$) then we refer to it as a "specialist" variant since it is specialized to have higher fitness than the wildtype in vaccinated host populations only. This categorization is useful because, for vaccine-adapted variants, generalists will increase in frequency and replace the wildtype regardless of the vaccine coverage whereas specialists require the vaccine coverage to be above a critical threshold before they will increase in frequency (**Figure 1**).

A second useful way to categorize a variant is to assess whether the absolute growth rate of infections that it causes is inhibited or facilitated by vaccination. The absolute growth rate of a vaccination-inhibited variant decreases as the vaccination coverage increases, whereas the absolute growth rate of a vaccination-facilitated variant increases with increasing vaccination (**Figure 1**). This categorization is useful because it speaks to whether the spread of infection will ultimately be lower or higher because of vaccination and subsequent vaccine-driven pathogen evolution. If a variant's growth rate is vaccination-inhibited, then increasing vaccination coverage will always reduce the overall spread of infection, even if the variant ultimately replaces the wildtype (**Figure 1a,c**). However, if a variant's growth rate is vaccination-facilitated, then if vaccination drives the variant to replace the wildtype it is possible that the overall spread of infection goes up (e.g., **Figure 1b**).

The categorization of variants in **Figure 1** is based on their per capita growth rates and therefore such plots are specific to epidemiological context. For example, early in an outbreak there is typically exponential growth in the number of infections, but as an outbreak progresses and/or non-pharmaceutical interventions (NPIs) are introduced, the force of infection will eventually decline, reducing all growth rates $r_{i,N}$ and $r_{i,V}$. Eventually, if the pathogen becomes endemic, the average growth rate across all variants will be zero. Notice, however, that the categorization of variants in **Figure 1** depends only on the *relative* growth rates and so the very same framework can be applied to any epidemiological context (e.g., in the early stages of an outbreak during exponential growth or at equilibrium once the pathogen is endemic). Moreover, if the relative ordering of variants does not change with epidemiological context, then their classification into one of the four categories will remain consistent regardless of what is happening epidemiologically (**Appendix 1**).

To conceptualize evolutionary change during a vaccination campaign we can then construct a plot of the absolute growth rate of different possible variants in each host type, locating on the plot each of the four types of variants from Figure 1 (**Figure 2**; alternative ways of plotting variants are discussed in Appendix 2). We can also use such a plot to illustrate how the nature of selection changes as a vaccination campaign proceeds. In Phase 1, when most hosts are naïve (i.e., *p* is small), selection will primarily favour variants with a larger growth rate in naïve hosts (**Figure 3a**). As we move to Phase 2 (**Figure 3b**), however, an increasing fraction of hosts are vaccinated (i.e., *p* increases) and selection shifts to primarily favouring variants with a larger growth rate in vaccinated hosts (**Figure 3c**). Throughout this transition the variants that appear can be specialists or generalists, and either vaccination-inhibited or vaccination-facilitated.

With this framework, evolutionary theory then makes some predictions about how we expect adaptation in novel pathogens to unfold during a vaccination campaign. As a pathogen adapts there will be occasional selective sweeps in which a new variant displaces the wildtype and becomes the new wildtype. The sequence of selective sweeps that occurs will be determined by both the direction of selection (the arrow in **Figure 3**) and the set of variants that happen to appear (**Box 2**). Initially, in a new host-pathogen association like SARS-COV-2, there will typically be abundant scope for adaptation to both naïve and vaccinated hosts and thus a great many of the variants that arise and become dominant will be generalist variants (**Figure 4a**). Over time, as the pathogen becomes better adapted to the novel host, and as vaccination coverage increases, there will be fewer new variants that increase fitness in both host types, leaving primarily specialist variants as the source of variation for further adaptation (**Figure 4b**). Thus, as a pathogen becomes increasingly adapted to a novel host, adaptation to vaccination will tend to result in the loss of some degree of adaptation to naïve hosts.

It is more difficult to make predictions about whether variants are likely to be vaccination-inhibited or vaccination-facilitated. At first one might wonder if vaccination-facilitated variants are even possible but, as we detail in the next section, such variants have been documented in some infectious diseases. Moreover, it is not difficult to imagine how such a variant might occur in SARS-CoV-2. For example, people with symptoms often isolate and socially distance so they do not infect others. A variant that evades vaccine-induced immunity in terms of its transmissibility could spread more rapidly in a fully vaccinated population than in a fully naïve population (i.e., it would be vaccination-facilitated) if vaccination reduces disease severity and so reduces the rate of isolation and social distancing. This hypothetical example helps to emphasize that categorizing a variant as vaccination-facilitated is solely a statement about its fitness (i.e., its ability to spread) and it carries with it no *a priori* implication about whether the spread of such a variant would ultimately lead to a greater or lesser amount of disease, either in an individual infection or in the population overall.

**Examples of Pathogen Adaptation to Vaccination**

Before considering examples of adaptation to vaccination it is important to stress that many vaccines have not been undermined by pathogen adaptation (e.g. smallpox, measles, polio). This lack of adaptation is hypothesized to result from two features commonly associated with vaccination [3]. First, because vaccination is a prophylactic intervention, it can keep pathogen numbers small within vaccinated hosts, which limits the generation and transmission of novel variants. Second, because vaccines typically induce immune responses against multiple targets on a pathogen, multiple genetic changes may be required to circumvent vaccine-mediated immunity [4]. Both features are expected to limit the ability of the pathogens to adapt to vaccination by hampering the accessibility of variants (fewer red dots in **Figure 3**, **Box 2**). However, for a handful of vaccines that do not keep pathogen densities below transmissible levels in the majority of infected hosts, or that do not induce immunity against multiple targets, evolutionary adaptation has occurred [3]. Given this, we look to these previous examples for guidance on possible outcomes of adaptation to vaccination in SARS-CoV-2.

The most direct way to determine how vaccines affect pathogen adaptation is through experimental evolution, yet we know of only one study that takes this approach. It involved a novel host-pathogen association of malaria parasites in laboratory mice [5]. Parasites were serially passaged for 20 generations through either vaccinated or naïve mice and allowed to evolve in response to these different treatments. The parasites became progressively better able to replicate in the host type they were evolving in, but they also evolved a better replication rate in the other host type as well. Moreover, vaccination inhibited the replication of all the evolved pathogens, demonstrating that the variants that arose during evolution were vaccination-inhibited generalists.

Most other data are observational and focus on pathogen species that have a longer association with their host. As expected from the earlier considerations, many vaccine-adapted variants appear to be specialist variants relative to the wildtype. For example, vaccine-adapted variants of hepatitis B virus arise that have altered surface antigens, making the vaccine less effective [6]. These variants cause sporadic breakthrough infections but they have not increased in overall number at the population level even as vaccination rates have increased [7, 8]. This suggests that, although they are more fit than the wildtype within vaccinated hosts, their spread from vaccinated hosts is apparently suppressed making them vaccination-inhibited specialists. In *Bordetella pertussis,* acellular vaccines that target PRN have led to the spread of vaccine-adapted variants that no longer express PRN [9]. These variants appear to be more fit than the wildtype in vaccinated hosts but less fit in naïve hosts making them specialist variants [10]. Variants also arise that overexpress the immunosuppressive PTX molecule, and these appear to be more fit than non-overexpressing variants in both naïve and acellular-vaccinated hosts [30]. Notably, fitness was not assayed in whole-cell vaccinated hosts limiting our ability to definitively classify the variants as specialists or generalists. In both sets of *B. pertussis* variants, however, the ability of these variants to spread in a vaccinated population appears to be less than in naïve populations [10, 11], making them all vaccination-inhibited variants.

Similar patterns often arise with vaccines used in farm animals, although the data necessary to distinguish between specialist and generalist variants are often inconclusive. For example, avian metapneumovirus vaccination suppresses virus shedding in turkeys, but less so for recent isolates of the virus than historical isolates, and no difference was detected between the isolates in non-vaccinated turkeys [12]. This difference has been credited to amino acid divergence in two genes [12]. Similarly, breakthrough against a vaccine for a fish bacterial pathogen *Yersinia ruckeri*, has been associated with a loss of the bacterial flagellum [13]. However, partial vaccine protection persists against all tested variants [14] again suggesting that these variants are vaccination-inhibited.

One strikingly different example is the chicken pathogen Marek's disease virus (MDV). MDV is an oncogenic virus that can cause paralysis and high levels of mortality [15], and a succession of vaccines have been developed and deployed in response to continual vaccine-driven evolution [16]. The vaccine-adapted variants that have been analyzed appear to be disfavoured in naïve chickens relative to the ancestral virus [17]. Nevertheless, unlike the examples described above, the vaccine-adapted variants of MDV transmit better from vaccinated chickens than from naive chickens [17]. These variants are therefore examples of vaccination-facilitated specialist variants. Notably, the overall prevalence of disease in the poultry industry was nevertheless reduced by vaccination despite this evolution [18] (as in **Figure 1d**).

Other examples of evolution in response to vaccination involve host-pathogen associations where multiple serotypes coexist and vaccines target only a subset of those serotypes. These situations are more complex because the very coexistence of serotypes suggests that multiple host types are present, possibly because of distinct immunological histories that have arisen through natural infection by the different serotypes. As a result, the framework in **Figures 2** and **3** would need to be extended with additional axes corresponding to the different kinds of hosts. Nevertheless, we can draw an analogy to the previous examples by viewing the set of serotypes targeted by the vaccine as the 'wildtype' and the non-targeted serotypes as the 'variants'. The fact that the wildtype and variant serotypes coexist suggests that, as expected, they are specialist variants. It is more difficult to categorize them as being vaccination-inhibited or vaccination-facilitated, but in all examples that we are aware of, the total prevalence of infection has either gone down or remained unchanged after the deployment of the vaccine. For example, vaccination against *Streptococcus pneumoniae* often resulted in no change in the total prevalence of bacterial carriage because non-targeted serotypes completely replaced vaccine-targeted serotypes following vaccination [19-21]. For human papillomavirus in contrast, vaccination reduced the total number of infections because non-targeted serotypes did not change in prevalence while vaccine-targeted serotypes became less common [22]. Other examples involving coexisting serotypes, including *Bordetella pertussis* [23], *Haemophilus influenzae* [24], *Neisseria meningitidis* [25], and rotavirus [26], appear to fall somewhere between these two extremes.

One final example is human influenza virus, which continually evolves in response to host immunity through a process known as antigenic drift, generating many sequential influenza variants over time [27]. To keep up with antigenic drift, flu vaccines are frequently updated. Again, this can be conceptualized in the current framework by introducing a new axis in **Figures 2** and **3** every time a new vaccine is introduced and/or a new immunological type of host arises. We were unable to find definitive data that addresses whether influenza variants tend to be generalists or specialists. Either way, existing data suggest that most novel variants arising through antigenic drift are partially inhibited by vaccination [28].

Thus, in the handful of cases where vaccine adaptation has been observed, specialist variants have been involved. This is consistent with our theoretical expectation that generalist variants will eventually give way to specialist variants as novel host-pathogen associations become more established (**Figure 4**). Moreover, most of those handful of cases involve vaccination-inhibited specialists. As a result, vaccination has generally resulted in a reduced overall spread of infection, even when vaccination drove the evolutionary advantage of the variants. We have identified examples of vaccination-facilitated specialist variants, but it is noteworthy that even in these cases it appears that such a vaccine-driven increase in the overall prevalence of infection has never been documented [29].

We are unaware of an example of a vaccination-facilitated generalist variant in any infectious disease. Such a variant would spread regardless of vaccine coverage, and it would also necessarily compromise our ability to control infection using that particular vaccine (**Figure 1c**). It is not clear if the apparent absence of such variants is because very few variants in this category are possible (**Box 2**), or if it is because generalist variants will be rare except when host-pathogen associations are new. As discussed above, it is possible to imagine such variants but again we stress that even if they arose their spread need not necessarily lead to a greater overall amount of disease in either infected individuals or at the population level.

## SARS-CoV-2

There is now substantial evidence that SARS-CoV-2 has been undergoing rapid adaptive evolution since its first appearance in humans. The first compelling data involved the spread of the Alpha and Delta variants because of their fitness advantages over the wildtype [30-32]. What does our framework tell us about the potential for SARS-CoV-2 adaptation to vaccination? Epidemiological data from several countries suggest that, as expected, the main vaccine-adapted variants to appear so far are vaccination-inhibited generalists (**Figure 5**). It is important to stress, however, that the evolutionary advantage of Delta has not been driven by vaccination even though it appeared in the vaccine era. The Delta variant increased in frequency in countries with very low vaccine coverage as well as in countries with relatively high vaccination coverage, suggesting it is a generalist. Data indicating that Delta is vaccination-inhibited are less direct and come both from epidemiological studies [33] and from neutralization assays [34]. Although these data only quantify one of the three components of fitness (see section '**The Relationship Between Pathogen Fitness and Infection Characteristics'** below), they show that while Delta is vaccine-adapted, current vaccines (BNT162b2 Pfizer-BioNTech, mRNA-1273 Moderna, and ChAdOx1 nCoV-19 Oxford-AstraZeneca) nevertheless still provide considerable levels of protection [35, 36]. The case for the Alpha variant being vaccine adapted is even less direct because Alpha spread and was then largely replaced by Delta before significant vaccine coverage existed in most countries. Thus, the epidemiological data clearly show that Alpha was advantageous relative to the wildtype in naïve hosts [32, 37, 38], but estimates of its fitness in vaccinated hosts again come from proxies using vaccine efficacy. The important point for both variants is that they would have become dominant regardless of whether vaccines had been deployed because they are generalists (**Figure 5**).

Although the above examples of evolution are not driven by vaccination, vaccine coverage is now reaching high enough levels in some countries that the possibility of vaccine adaptation has become a real concern. As mentioned earlier, vaccine-driven evolution has tended to occur in other pathogens when either the benefits of prophylaxis are small (e.g., the vaccine does not sufficiently suppress viral replication below transmissible levels) or when they target a small number of viral epitopes [3, 39]. Data increasingly suggest that at least the first of these is true for SARS-CoV-2 [40-43]. As SARS-CoV-2 adapts further to humans we might therefore expect that specialist variants will begin to appear that have even higher reproductive success in vaccinated populations but where this increased adaptation to the vaccine comes at a cost of reduced reproductive success in naïve populations. Indeed, at the time of writing yet another new SARS-CoV-2 variant of concern (labelled Omicron) appears to be undergoing a selective sweep in some parts of the world. Current estimates suggest that its selection coefficient is quite large, between 0.2/day and 0.4/day [44]. Furthermore, growing evidence suggests that Omicron is less well controlled by current vaccine schedules than is the Delta variant but it is still too early to categorize Omicron definitively (**Figure 5**).

So far as we know, vaccination-facilitated variants in SARS-COV-2 have not yet been reported and, depending on the available genetic variation (**Box 2),** it is possible that they never will arise. For a variant to be vaccination-facilitated, the vaccine would have to either increase the rate at which the variant generates new infections and/or decrease the rate at which existing infections caused by the variant are lost from circulation through recovery, isolation, or death. In principle, molecular processes involving antibody dependent enhancement of cell infectivity (ADE) could provide a mechanism by which vaccine facilitation occurs [34, 45, 46], but we know of no evidence that ADE has increased transmission in any infectious disease. Vaccination could also increase the rate at which a variant generates new infections if vaccinated people engage in more risky behavior (e.g. are allowed entry to concerts and bars [47]). The other type of variants that could be facilitated by vaccination are variants whose transmission is curtailed because of the disease severity that they cause (e.g. leading to isolation). Vaccination, which is aimed at reducing disease severity, could facilitate silent or semi-silent spread of such variants (**Box 2**) in a manner directly analogous to the variants facilitated by the first generation vaccines against Marek's disease [17]).

In the longer term, if variants like those hypothesized above appear and spread, thereby compromising the utility of the vaccine, it is likely that boosts and new vaccines would be introduced. Furthermore, as SARS-CoV-2 spreads in the human population and presumably becomes an endemic virus, the number of people with an immunological history due to natural infection will increase significantly as well. In both cases, the framework presented here will need to be extended to account for multiple host types. Making longer-term predictions for such cases is difficult at this stage because a great deal will depend on the nature of the genetic variation that is possible (**Box 2**).

**The Relationship Between Pathogen Fitness and Infection Characteristics**

The above analysis focuses solely on pathogen fitness. One thing missing from this discussion is a consideration of how vaccination might drive the evolution of infection characteristics like vaccine efficacy or disease severity. To better illustrate the relationship between the fitness of a variant (as measured by the growth rate of infections that it causes) and the characteristics of the infection, we can decompose the absolute growth rate $r_i$, of a variant into three main components of fitness (**Box 1 and 3**): (i) *infectivity* - the probability that, upon exposure, a variant infects either type of host; (ii) *transmissibility* - the rate at which a variant produces infectious propagules that contact uninfected individuals; and (iii) *infection duration* – how long a variant produces infectious propagules in either type of host before the infectious period ends through recovery, isolation, or death. All else equal, variants with increased infectivity, increased transmissibility, or increased duration of infection will have an increased growth rate.

*Vaccine efficacy against infection* - The infectivity of a variant is a key property for determining how well a vaccine works against a variant. If $\sigma_N$ and $\sigma_V$ denote the infectivity of a variant in naïve and vaccinated hosts respectively, then vaccine efficacy (VE) is the proportional reduction in infectivity that vaccination confers, given by $VE = 1 - \sigma_V/\sigma_N$. This highlights two important things about the utility of VE for understanding the evolutionary epidemiology of vaccine-adapted variants. First, because *VE* is a measure of the relative infectivity of a variant in vaccinated versus non-vaccinated hosts, a variant can have a reduced *VE* as a result of an increase in $\sigma_V$ and/or a decrease in $\sigma_N$. Second, *VE* involves only one of the three different components of fitness and so it provides only partial information for determining the fate of a variant or the consequences it will have if it sweeps to fixation. For example, the Beta and Gamma variants of SARS-Cov-2 both appear to reduce VE [48] yet, to date, neither has become the dominant variant. Measures of *VE* that capture other components of pathogen adaptation to vaccinated hosts do exist [49].

A related issue arises in discussions of vaccination that center around so-called "escape variants". Although this term is not always defined precisely, it is often used in reference to variants that differ in epitope and so are able to escape a specific immune response as measured in inhibition assays *in vitro* [48, 50-53]. For example, SARS-CoV-2 variants are sometimes characterized by both their transmissibility (as measured by their overall growth rate and/or $R_0$) and their performance in inhibition assays. We have purposefully avoided doing so here because this approach conflates the mechanism through which a variant is potentially adapted to vaccinated hosts (i.e., escape from a specific immunity and so greater ability to replicate within an individual) with the source of selection that favours the variant (e.g., increased infectivity). It is useful to keep these notions distinct because there are many different mechanisms through which a variant can be adapted to vaccinated hosts **(Box 1)** and each of these can affect any of the three main epidemiological components of fitness (i.e., infectivity, transmissibility, infection duration). Therefore, we believe the most consistent, general, and agnostic way to characterize variants is as described in **Figure 2.** Ideally, we would also quantify multiple infection characteristics (infectivity, transmissibility, and infection duration) for variants that arise, along with this quantification of fitness (**Appendix 2**). Such an approach is possible for SARS-CoV-2 using the unprecedented availability of genetically resolved, real time epidemiological data (**Box 3** and **Figure 5).**

*Disease Severity* – Arguably the most important infection characteristic from the standpoint of human health is the severity of disease caused by a variant. Most definitions of severity capture both the morbidity and the mortality caused by infection. As such, severity can affect all three components of fitness. For example, high disease severity might reduce infection duration through increased mortality, or it might reduce the transmissibility through a reduction in activity level and thus the contact rate of infected individuals [54]. In most cases disease severity *per se* is disadvantageous to the pathogen and thus selected against [55]. It is nevertheless difficult to make predictions about how disease severity will evolve because variants that cause more severe disease might have increased fitness relative to the wildtype through differences in other components of fitness [56]. For example, data suggests that the Alpha variant of SARS-CoV-2 may cause more severe disease than the Wuhan wildtype [57, 58], but it nevertheless has higher fitness because its transmissibility is higher. Also, severity of the disease may be partially mediated by the host immune response and recent studies suggest that some antibodies may "enhance" the replication of the virus and may induce more symptoms [45, 59]. A SARS-CoV-2 variant that could escape from neutralizing antibodies and exploit this enhancing effect could lead to greater disease severity in vaccinated or previously infected hosts [46]. This illustrates that, although we can make quite robust and reliable predictions about the evolution of pathogen fitness in naïve and vaccinated hosts, it is harder to make predictions about the underlying components of fitness or disease severity since variants with very different values of the three fitness components can nevertheless have the same overall fitness (**Box 1** and **3**). This means that pattens of evolution in these infection characteristics are likely to be somewhat idiosyncratic. This is a major reason why we cannot extrapolate the evolutionary trajectories of such traits from one pathogen to another.

Despite the lack of robust theoretical predictions about disease severity, a few observations from other infectious diseases could be relevant to SARS-CoV-2. First, vaccine protection tends to be even more evolutionarily robust against disease than against infection. This conclusion arises from the observation that when pathogens have evolved in response to vaccines in the past, vaccinated individuals that are infected by a pathogen tend to have better outcomes than non-vaccinated individuals [29]. A potential concern is if there are enhancing effects of antibodies on disease severity [60, 61], as there may be for COVID [45, 46, 59]. Second, for pathogens with coexisting serotypes, vaccine-driven serotype replacement could in principle increase or decrease overall disease burdens if different serotypes have different propensities for causing disease, as they often do (for example, [62]). Rational design of variant-based vaccines must therefore consider both the current prevalence of each variant and their likelihood of causing disease given infection. Third, under certain conditions, vaccines may lead to the evolution of highly virulent variants. The best example of this is MDV in which highly virulent variants of the virus kill their hosts so quickly that they are unable to persist in the absence of vaccination [17]. Vaccines ameliorate disease severity of MDV and so they allow hosts infected by these highly virulent variants to remain alive but they do not prevent transmission. Despite this effect, however, vaccinated chickens exposed to these highly virulent variants are nevertheless better off than non-vaccinated chickens exposed to the original wildtype. On the other hand, non-vaccinated chickens are now at greater risk of infection with variants causing more severe Marek's disease than they were prior to the introduction of the vaccine. Regardless of whether SARS-CoV-2 follows this path (**Figure 5**), vaccination remains our most effective tool to mitigate the epidemic, as was the case with MDV [63]. Vaccination also reduces the number of cases which may also slow down the flux of new mutations and thus the probability of viral adaptation (**Box 2**).

**Implications**

If further adaptation of SARS-CoV-2 occurs in response to vaccination, then our framework and the examination of previous experimental and empirical examples suggest that the long-term outcome will likely yield specialist variants. The path to getting there will likely involve vaccination-inhibited variants meaning that we are likely to, at least partially, retain the benefits of vaccination with first-generation COVID vaccines in the short term. In the meantime, there is an urgent need to monitor the epidemiology and evolution of the virus [39]. This will better characterize newly arising variants (**Box 3**) and make it possible to decide if, like for flu, new vaccines are needed to counteract viral adaptation.

It is also critical to stress that concerns about possible future viral evolution are not a reason to withhold currently available vaccines. First, vaccines are currently greatly reducing disease burdens and saving lives [64]. Second, as discussed above, much of the evolution currently occurring in SARS-CoV-2 involves generalist variants and so would be occurring regardless of whether we deployed existing vaccines. Third, immunity arising from natural infections will also impact on-going viral evolution. Currently, it is impossible to know whether natural immunity or vaccine-induced immunity will be the stronger evolutionary driver. Fourth, even with the Delta variant, current mRNA vaccines substantially reduce the probability of infection and infection duration compared to infections in naïve individuals [42, 43, 48]. That itself very substantially reduces evolutionary potential (**Box 2**).

Going forward, it is quite possible that new vaccine schedules (e.g., higher doses, boosters, combinations of existing vaccines) or next-generation vaccines (e.g., new RNA sequences, mucosal vaccines) will be required to deal with SARS-CoV-2 evolution. A diversity of vaccine types are already being used around the globe, and vaccine schedules in many locations are being continually adjusted. If this diversity generates relevant immunological heterogeneity within and among populations, then natural selection could favor different viral variants at different times in different locations, and perhaps even result in the coexistence of several variants. If so, vaccination programs may need to be continually adjusted at a national or regional level, as is necessary to control coronaviruses in agriculture [65, 66]. The more that vaccination suppresses transmission, targets multiple epitopes, and more effectively inhibits infection and within-host replication and so mutation and recombination, the better it will be at slowing the rate of adaptation (**Box 2**) and providing sustainable long-term efficacy [39].

## Summary

In the history of human and animal vaccination, there are few documented cases of vaccine-driven evolution. For situations where adaptation to vaccines occurs, either through the vaccination itself or through immunity due to natural infection, we propose a typology of vaccine-adapted variants based on their fitness in naïve and vaccinated host populations (**Figure 1**).

Adaptation occurs when a novel variant is more fit than its predecessors. The fitness of a variant is measured by its per-capita growth rate of the number of infections that it causes (i.e., the number of new infections per infection per unit time).

In the early phase of pandemics, we expect the rise of variants that are better at spreading than their ancestors in both naïve and immunized hosts (generalists). Later, viral evolution should involve specialised adaptations to immunized hosts, and so some decrease of adaptation to naïve hosts.

Both generalist and specialist variants can be inhibited by vaccination, where the growth rate of infections decrease as vaccine coverage increases. Under these circumstances, even if the impact of vaccination is eroded by viral evolution, the overall spread of infection is still reduced by vaccination.

Vaccination-facilitated variants can arise. In this case, the overall spread of infection could theoretically go up as vaccination rates increase but this does not imply that the overall level of disease necessarily will increase in either an individual infection or in the population overall.

Evolution is more than selection. Our framework predicts the direction and strength of selection, but does not precisely predict the evolutionary trajectory that will be followed because there is no way of knowing in advance what phenotypes are available to the virus genetically (via mutation or recombination) (**Box 2**).

Evolution is more than mutation. There is no way of knowing in advance how particular mutations relate to the multiple dimensions of the fitness landscape, even if they may have an advantage on a particular dimension in a laboratory assay (**Box 1**).

So far, the SARS-CoV-2 variants of concern that have become dominant have been vaccination-inhibited generalists that would have spread regardless of vaccination. We expect more generalist variants to arise and spread until the mutational supply has been exhausted. Once exhausted, we expect further adaptation to result from the spread of specialist variants. Whether these variants will be vaccination-inhibited or vaccination-facilitated will depend on mutational availability.

Beyond those expectations, a priori prediction about future vaccine efficacy and disease severity for SARS-CoV-2 is not possible. Molecular epidemiological surveillance will be critical for detecting and characterizing viral adaptation as it unfolds (**Box 3**).

**Acknowledgements:** We thank the Newton Institute, the RAMP continuity network and the JUNIPER consortium for organising a workshop on Evolutionary Implications of the COVID-19 Vaccination Programme in spring 2021 which sparked the present work.

**Funding:** TD was supported by a grant from the Natural Sciences and Engineering Research Council of Canada. DK & AR were supported by Institute of General Medical Sciences, National Institutes of Health (R01GM140459, R01GM105244 respectively) and the UK Biotechnology and Biological Sciences Research Council as part of the joint NSF-NIH-USDA Ecology and Evolution of Infectious Diseases program. DK was also supported by National Science Foundation grant DEB-1754692. SG was supported by the CNRS and by ANR-17-CE35-0012 from the Agence Nationale de la Recherche. The funders had no role in study design, data collection and analysis, decision to publish, or preparation of this article.

**Figure Captions**

**Figure 1: Four types of vaccine-adapted variants.** Solid lines depict the growth rate of the population of infected individuals for the wildtype (blue) and for a variant (red) as a function of vaccination coverage. Vaccination decreases the growth rate of the wildtype ($r_N > r_V$). Quantities $\Delta r_N$ and $\Delta r_V$ are the differences in growth rate between the variant and the wildtype in naïve and vaccinated hosts, respectively. Colored shading indicates which type prevails evolutionarily: the wildtype (light blue shading) or the variant (light red shading). Panels **(a)** and **(b)** are generalists - the variant is also better adapted to naive hosts ($\Delta r_N > 0$). Generalist variants will outcompete the wildtype even in the absence of vaccination. Panels **(c)** and **(d)** are specialists – the variant is maladapted to naive hosts ($\Delta r_N < 0$). Specialist variants will outcompete the wildtype only above a critical vaccination threshold. Panels **(a)** and **(c)** are vaccination-inhibited variants - the growth rate of the variant decreases with increasing vaccination. As a result, the growth rate of infections after adaptation (i.e., after fixation of the fittest type) in a fully vaccinated population (black dot) is always lower than that in a fully naïve population (white dot & dashed line). Panels (**b**) and (**d**) are vaccination-facilitated variants - the growth rate of the variant increases with increasing vaccination. As a result, the growth rate of infections after adaptation in a fully vaccinated population (black dot) is always higher than that in a fully naïve population (white dot) for generalist variants (panel **(c)**) but it can go either way for specialists (panel **(d)** - only the case where it is lower is shown).

**Figure 2: Four types of vaccine-adapted variants.** A plot of the growth rate of variants in a fully naïve, $r_{i,N}$, and a fully vaccinated, $r_{i,V}$, population. Blue dot indicates location of the wildtype. Uncolored region corresponds to variants whose growth rate in vaccinated hosts is less than that of the wildtype and so are vaccine-maladapted (and so ignored in our discussion). Different coloured regions correspond to the 4 types of variants from **Figure 1** (labels **(a)**-**(d)** correspond to panels **(a)**-**(d)** from **Figure 1**). Additional types of variants are presented in **Figure S1**. See Appendix 2 for a discussion of alternative ways to visualize variants.

**Figure 3: Selection and genetic variation.** A plot of the growth rate of all viable variants i in a fully naïve and a fully vaccinated population (black dots). Large blue dot denotes the current wildtype. Red dots are those variants that are most accessible from the wildtype (see **Box 2**). Note that the location of all variants along the $r_{i,V}$ axis is specific to a vaccine and will be different for different vaccines. All variants in white region are selectively advantageous but variants in the direction of the selection arrow are most strongly favoured (dashed lines indicate contours of overall growth rate). Variants in the grey region are disfavoured by selection. The direction of selection arrow is upwards in a fully naïve population ($p = 0$) (panel **(a)**) and shifts towards the right as the level of vaccination (and/or fraction of hosts with exposure to the wildtype through natural infection) increases (panels **(b)** and **(c)**).

**Figure 4: Pathogen adaptation during vaccination.** A plot of the growth rate of all viable variants in a fully naïve and a fully vaccinated population (dots). Large blue

dot denotes the phenotype of the current wildtype and black arrow indicates direction of selection (i.e., the variants that are most advantageous). Variants in the grey region are disadvantageous. Note that the location of all variants along the $r_{i,V}$ axis is specific to a vaccine and will be different for different vaccines. Coloured regions indicate the four different kinds of variants. **(a)** Early in a novel host-pathogen association (and in the first phase of the vaccination campaign). Many potential new variants will be better adapted to both host types (i.e., they will be generalists). **(b)** Later in the association, when the pathogen is better adapted to its novel host (and vaccination levels are higher). The evolutionary trajectory of successive fixation events leading to the new wildtype variant is indicated with the succession of blue dots. Note how the change in the location of the blue dot can affect the typology of some variants (i.e., a variant that was identified as a generalist in the early stage of adaption could later become a specialist relative to the more recent form of the virus). Once the level of adaptation is high (panel **b**) most advantageous variants that appear will tend to be specialists. Even though generalists are still more strongly favoured by selection there are fewer of them that can arise.

**Figure 5: Graphical representation of SARS-CoV-2 adaptation to naïve and vaccinated hosts.** Blue dot denotes location of the Wuhan wildtype, which is relatively poorly adapted to both naïve and vaccinated hosts. The bars give the range of plausible values for the growth rate of each variant in fully naïve or fully vaccinated hosts (where fully vaccinated means two doses of current mRNA vaccines). The Wuhan wildtype and the Alpha variant are vaccine-maladapted relative to Delta variant (see **Figure 2**). Note the dashed black lines indicate where the growth rates are zero (the three variants are able to grow in a fully naïve population but not in a fully vaccinated population). The red ellipse indicates the likely position of the newly emerged Omicron variant at the time of writing (early January 2022). See **Appendix 2** for details regarding the construction of this figure.

**Box 1 – Mechanisms of vaccine adaptation**

Our focus is on the ability of a variant to spread between hosts, and this ability can arise from several different mechanisms operating within an infected individual. Pathogens have evolved a vast diversity of countermeasures against natural immunity, many of which will also be highly effective against vaccine-induced immunity and so are expected to be involved in adaptation to vaccination. The list below is intended to be illustrative of the diversity of possible within-host adaptations, rather than comprehensive.

**Immune evasion** (avoiding anti-pathogen responses).
- Antigenic change.
- Antigenic loss. Inactivation or deletion of molecules targeted by host responses. Examples include loss of toxins (diphtheria, pertussis).
- Antigenic repertoires. Changes in genes controlling the rates at which pathogens generate and expose novel antigens (e.g. trypanosomes, malaria).
- Increased cell-cell infection to evade antiviral humoral immunity which threatens cell-free infection [53].
- Altered tissue tropism to immune-privileged sites.

**Immune suppression** (dampening or mis-directing anti-pathogen responses).
- Up-regulation of enzymes to degrade effector molecules (e.g., ptxP3 in pertussis)
- Production of immune-regulatory molecules such as cytokine mimics (e.g. pox viruses) and immune antagonists (e.g. Orf9b and Orf6 in Alpha variant of SARS-CoV-2, [67]).
- Production of substances that drive inappropriate responses (e.g. helminths)
- Production of 'smoke screen' molecules, which distract immune effector molecules (e.g. malaria, [68])

**Immune exploitation** (utilizing host responses)
- Antibody-dependent enhancement (e.g. [45, 46])

**Life-history mediated countermeasures against immunity**

Direct countermeasures against immunity, such as those listed above, are not the only possible within-host mechanisms of vaccine adaptation. A very different suite of potential mechanisms has to do with where, when, and how fast pathogens replicate.
- Variants that replicate earlier or faster can overwhelm the immune response, at least initially.
- Variants that replicate more slowly can potentially remain below immune detection for longer (e.g., many chronic viral infections).
- Variants which can exploit altered host cell invasion pathways can have an advantage when primary pathways are blocked by host immunity.

Traits underpinning these mechanisms can include higher binding affinity to host receptors, large burst sizes (number of pathogen progeny released from a host cell), altered latency (dormancy in host cell) and changes in the investment of within-host replication relative to transmission stage production (e.g., malaria).

Finally, where transmission is restricted by disease severity (for instance, via host death or hospitalization), vaccination, can enhance pathogen transmission by reducing disease severity (e.g., Marek's disease).

Most of the traits listed above can be studied in a variety of *in vitro* and *in vivo* models, with native pathogens or novel expression systems like pseudoviruses. Often, *in vivo* studies are also possible, using animal models and, in some cases, human subjects. In most cases, it is very challenging to link within-host mechanisms to between-host fitness because individual traits are often, at best, correlates or partial determinants of one or more of infectivity, transmissibility and infection duration (which are the three key components of fitness). Fitness *per se* (i.e., the growth rate of the number of infections) and other components of fitness can also be inferred in real time from rates at which the different variants spread in the human population (**Box 3**).

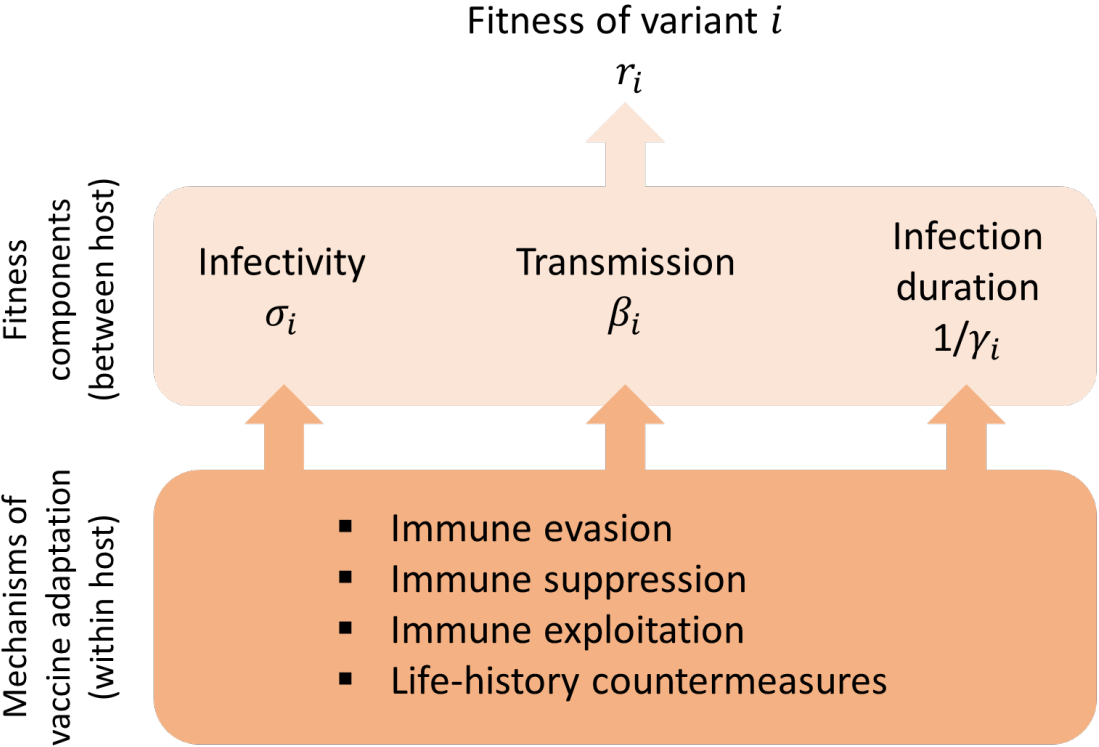

**Figure Box 1: The fate of a variant i within a host population is determined by three key components of fitness, each of which can be affected by several within-host mechanisms of adaptation.** All else equal, variants with increased infectivity, increased transmissibility, or increased duration of infection will have an increased fitness (rate of spread in a population). As indicated in equation (1) fitness depends on both the amount of adaptation to naïve and vaccinated hosts. Within-host processes impact those three components of fitness to varying extents and, in turn, individual viral mutations can affect those processes to varying extent. Some within-host mechanisms of adaptation can be measured directly in in vitro assays. Some components of viral fitness can be inferred from evolutionary epidemiological studies (**Box 3**).

**Box 2 – Mutation and adaptation to vaccination**

Pathogen adaptation requires variation in fitness among variants. New variants arise from mutation during replication and from recombination when distinct variants coinfect the same host. It is important to distinguish between the rate at which new variants arise and how their fitness differs from the wildtype.

**The rate at which variants arise**

Mutations are continuously generated during the replication of the virus within infected hosts. The rate at which this occurs is proportional to the rate at which genomic changes occur during replication, and the amount of replication that is taking place. Vaccination reduces the amount of replication taking place in two ways. First, at the within-host level, if a vaccinated host is infected, a vaccine-primed immune response is expected to reduce the viral load and to clear the infection faster. Second, at the between-host level, the rollout of vaccination is expected to reduce the number of infected hosts (both naïve and vaccinated). These effects are tempered for imperfect (or leaky) vaccines, however, because they have a lower ability to reduce pathogen replication and to prevent infection.

**The fitness effects of variants**

The fate of a new variant is determined by how the rate of change of number of infections it causes differs from that of the wildtype in both naïve and vaccinated populations (i.e., where it falls in **Figure 3** relative to the wildtype). To this end it is useful to distinguish between the set of variants that are *possible* (all the dots in **Figure 3**) and the set of variants that are easily *accessible* from the wildtype (the subset of red dots in **Figure 3**). There will be biological constraints on the magnitude of growth rate that is possible in the two host types and therefore all the dots in **Figure 3** will fall within some specific region of the plane. Most mutations are expected to be deleterious or have little effect, but some may result in a larger growth rate than the wildtype [69, 70]. Hence, we expect a high density of possible phenotypes (black dots in **Figure 3**) with low fitness relative to the density of phenotypes that increase fitness in both host types. Within this set of possible variants, some will be more readily accessible from the current wildtype than others for several reasons. First, some variants might be multiple mutational or recombinational steps away from the wildtype and so will be exceedingly unlikely to arise. For example, the lack of adaptation of measles virus to vaccines despite decades of global vaccination is potentially because variants that can escape a polyclonal antibody response require at least five new mutations to the H glycoprotein [4]. Second, competition between the variant and the wildtype within an infection can promote (or hamper) the variant's ability to reach a density high enough for onward transmission to occur. For example, in novel host-pathogen associations, mutations that are beneficial for within-host competition are also likely to be beneficial in other respects, including their ability to spread at the between-host level simply because more generalist variants are accessible when the wildtype is poorly adapted to its host (see **Figure 4a** but when axes are within- and between-host fitness). As the association becomes more established, however, variants that are successful within hosts will tend to have reduced success at the between-host level. This effect of within-host selection biasing the set of variants that are accessible to between-host selection is likely also modulated by the leakiness of the vaccine [71].

**Vaccination and the speed of pathogen adaptation**

Faster rollout and more effective vaccines will, all else equal, limit the emergence of new variants. Hence, the use of leaky vaccines (and the occurrence of chronic infections in immunocompromised hosts) could speed up pathogen adaptation both because they increase the flux of mutation relative to the use of non-leaky vaccines and because they facilitate the within-host rise of some vaccine-adapted variants. Once a vaccine-adapted variant is circulating in the population, the influence on evolutionary adaptation of the rate at which it arises through mutation is negligible compared to the selection acting on the variant (e.g. the dynamics of the Alpha variant of SARS-CoV-2 at the end of 2020 in UK was driven by selection, not by the flux of mutations). In this case, the speed of pathogen adaptation is mainly driven by selection and different targeted vaccination strategies may provide ways to slow down this adaptation [72-74].

**Box 3 – How to characterise the fitness of SARS-CoV-2 variants?**

The ongoing pandemic of SARS-CoV-2 is characterised by an unprecedented access to incidence and sequencing data in real time. This data provides a unique opportunity for quantifying the underlying components of viral fitness (infectivity, transmissibility, and infection duration) related to adaptation to naïve and vaccinated hosts. Three main dynamical variables carry useful information about these components of fitness (**Appendix 1**).

First, the per capita growth rate of the epidemic during vaccination provides information about the potential emergence and the spread of new variants. Any deviation from the predicted drop in incidence of the wildtype due to increasing vaccination coverage could signal the spread of a vaccine-adapted variant ($\Delta r_V > 0$).

Second, analysis of the change in frequency of a variant allows some inference to be made about which components of fitness underly adaptation to vaccination. We show in **Appendix 1** that the magnitude of change in the frequency of a variant will be proportional to the availability of susceptible hosts if the variant obtains its advantage through increased transmissibility, $\beta$, or infectivity, $\sigma$, but this change will be independent of susceptible hosts if the variant obtains its advantage through a longer infection duration. Therefore, as the availability to susceptible hosts varies with lockdowns and other NPIs, tracking how this affects the change in variant frequency can inform us about the mechanism underlying the variant's success [56, 75].

Third, the over representation of a variant in vaccinated hosts can be used as an early signal that the variant is adapted to the vaccine. We show in **Appendix 1** that the difference in variant frequency between naïve and vaccinated hosts (i.e., the genetic differentiation of the viral populations in the two types of hosts) is mainly governed by the relative infectivity of the variant in vaccinated hosts, but not by its transmissibility. Hence, the analysis of these three dynamical variables provides a way to begin disentangling the three major components of fitness.

Figure 1

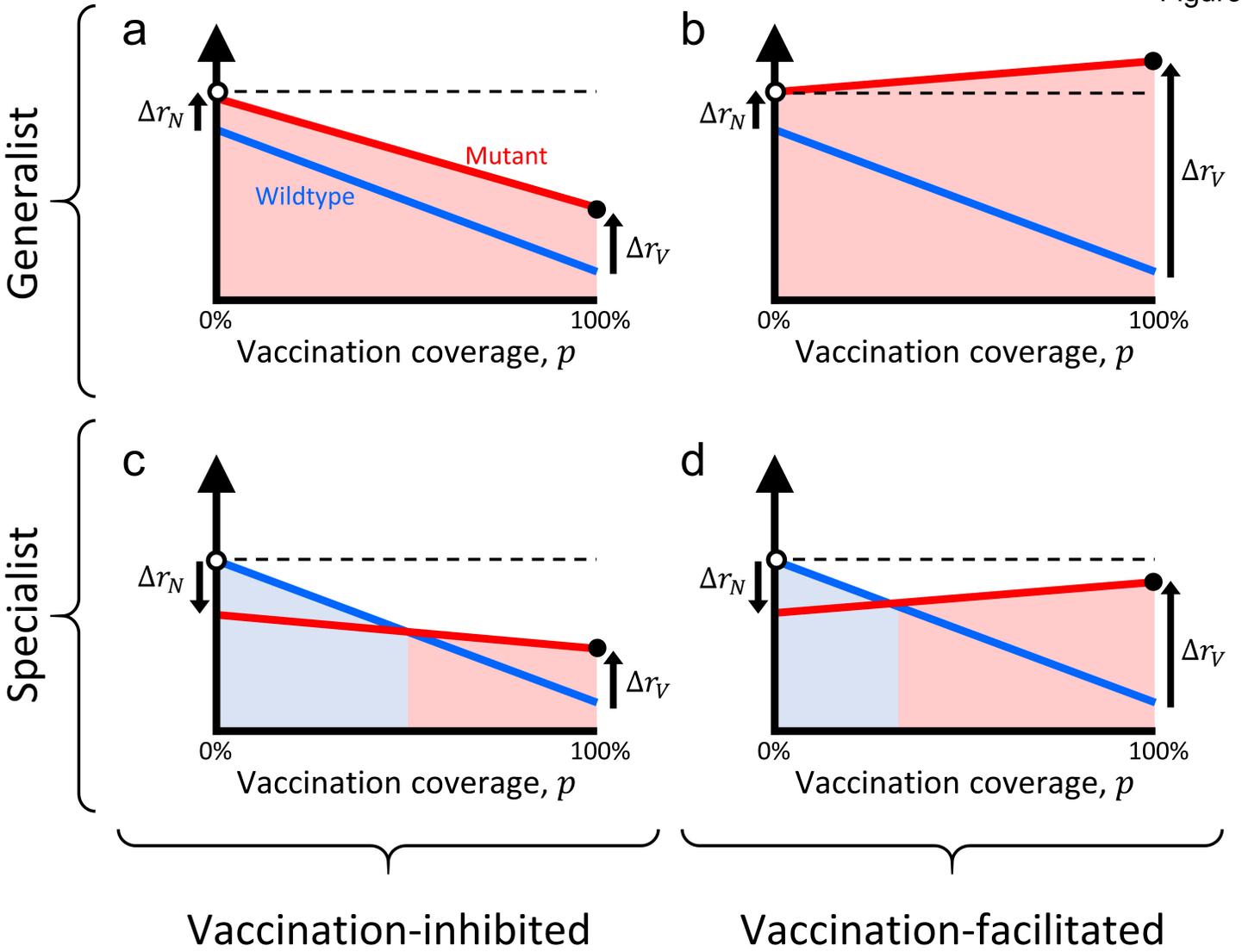

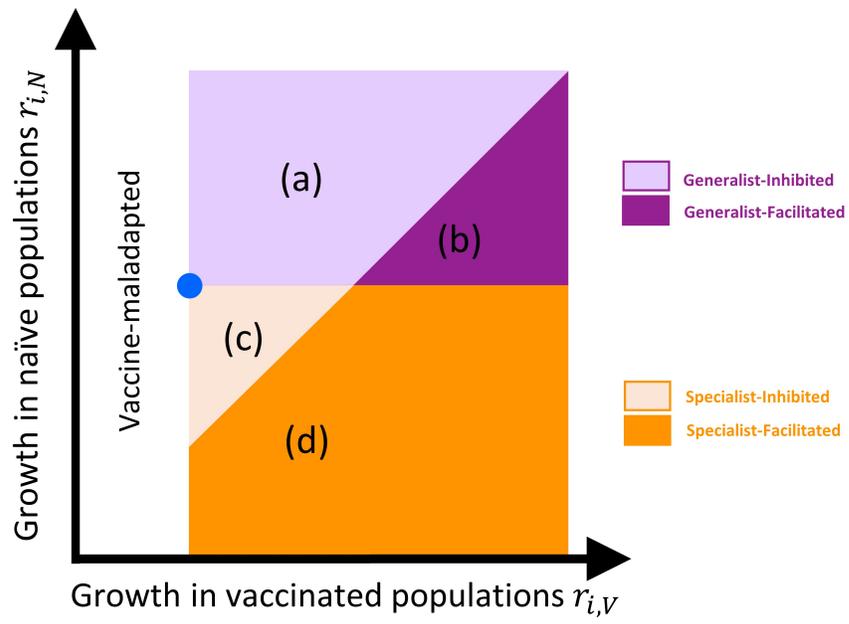

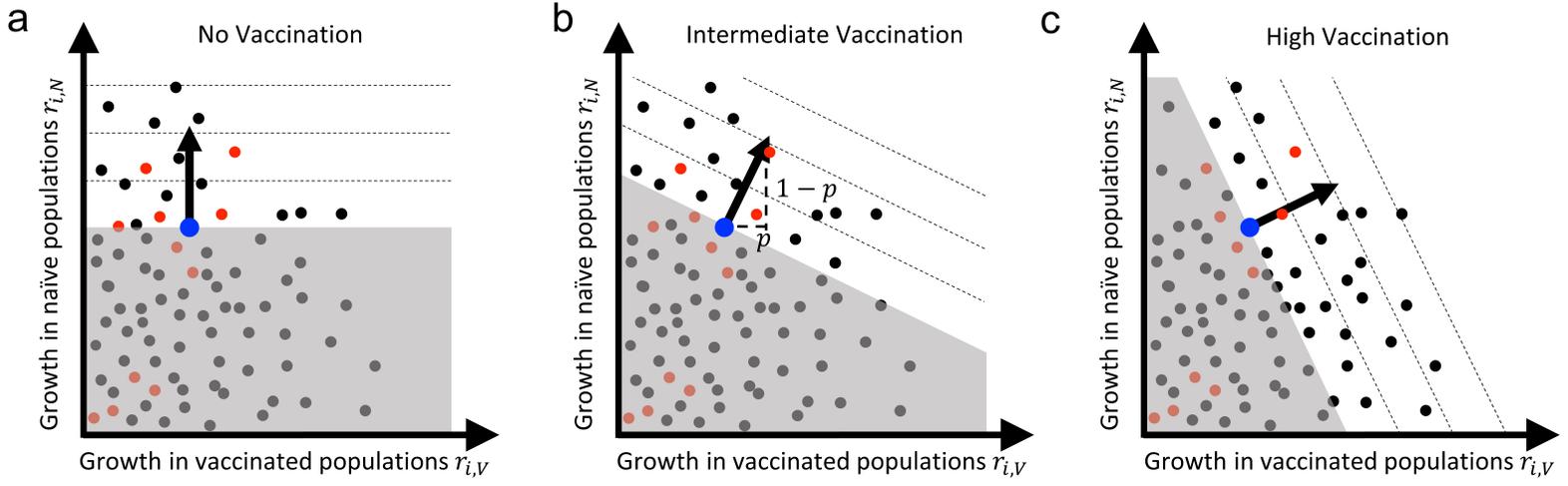

a    No Vaccination

Growth in naïve populations $r_{i,N}$

Growth in vaccinated populations $r_{i,V}$

b    Intermediate Vaccination

Growth in naïve populations $r_{i,N}$

$1-p$

$p$

Growth in vaccinated populations $r_{i,V}$

c    High Vaccination

Growth in naïve populations $r_{i,N}$

Growth in vaccinated populations $r_{i,V}$

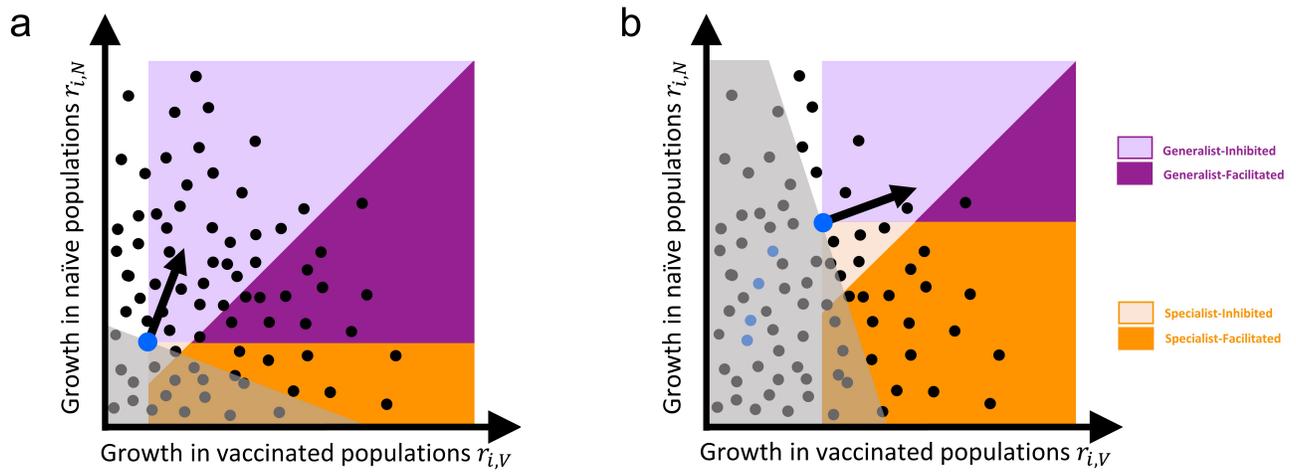

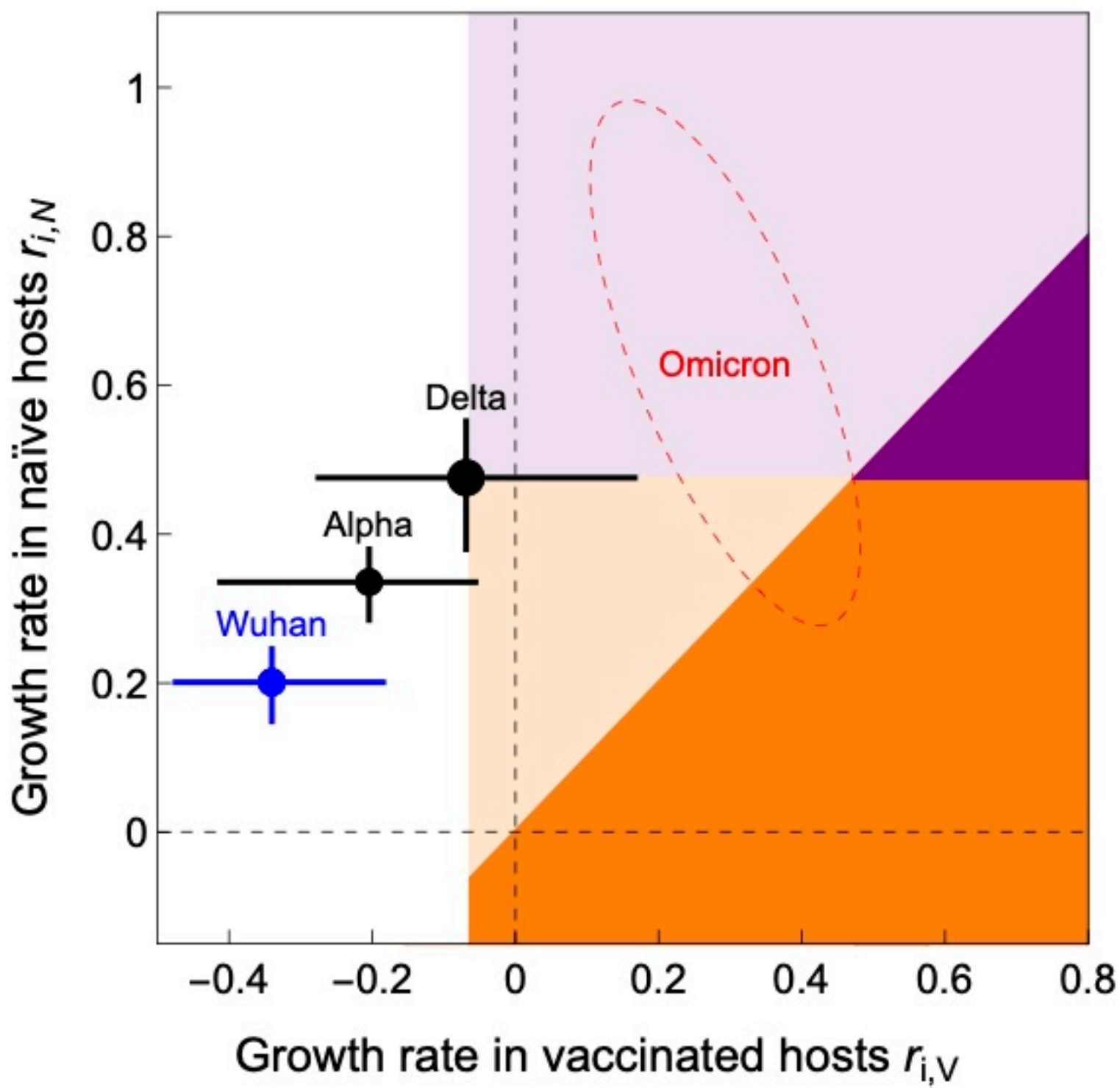

# <u>Supplementary Information</u>

# Appendix 1

In this appendix we derive the expressions presented in the main text, and we also show how changes in the three main components of fitness of a variant affect the evolutionary dynamics.

## 1. The model

We track the dynamics of variant $i$ in a host population with a density $S_N$ of naïve hosts and a density $S_V$ of vaccinated hosts using the following system of differential equations:

$$\dot{I}_i^N = h_i \sigma_i^N S_N - \gamma_i^N I_i^N$$
$$\dot{I}_i^V = h_i \sigma_i^V S_V - \gamma_i^V I_i^N$$
$$h_i = \beta_i^N I_i^N + \beta_i^V I_i^V$$

The reproductive success of a variant is determined by three components of fitness:

- $\beta_i^N$ and $\beta_i^V$: the transmission rate of variant $i$ from naïve and vaccinated hosts
- $\sigma_i^N$ and $\sigma_i^V$: the infectivity of variant $i$ in naïve and vaccinated hosts
- $\gamma_i^N$ and $\gamma_i^V$: the infection duration of variant $i$ in naïve and vaccinated hosts

Below we present the derivation of the three dynamical variables mentioned in Box 3 that capture the epidemiological and evolutionary dynamics during adaptation to vaccination (equations (S1), (S2) and (S3) below).

## 2. The growth rate of the epidemic

If the change in the density of susceptible hosts occurs slowly relative to the spread of infection then the per capita growth rate of infections will be given by the dominant eigenvalue $r_i$ of the matrix:

$$\mathbf{R}_i = \begin{pmatrix} r_i^{NN} & r_i^{VN} \\ r_i^{NV} & r_i^{VV} \end{pmatrix}$$

where

$$r_i^{NN} = \beta_i^N \sigma_i^N S_N - \gamma_i^N$$
$$r_i^{VN} = \beta_i^V \sigma_i^N S_N$$
$$r_i^{VV} = \beta_i^V \sigma_i^V S_V - \gamma_i^V$$
$$r_i^{NV} = \beta_i^N \sigma_i^V S_V$$

If we further define $\delta_i = \gamma_i^V - \gamma_i^N$ then we can write

$$r_i = (1-p)r_{i,N} + pr_{i,V} - \frac{S_V \beta_i^V \sigma_i^V}{S_N \beta_i^N \sigma_i^N + S_V \beta_i^V \sigma_i^V} \delta_i + O[\delta_i]^2$$

with: $r_{i,N} = S\beta_i^N \sigma_i^N - \gamma_i^N$, $r_{i,V} = S\beta_i^V \sigma_i^V - \gamma_i^V$, and where $S = S_N + S_V$ and $p = \frac{S_V}{S_N + S_V}$ is the *coverage of vaccination* (i.e., the fraction of the uninfected population that is vaccinated).

When $\delta_i = 0$ this simplifies to:

$$r_i = (1-p)r_{i,N} + pr_{i,V}$$

which is equation (1) of the main text. Notice, however, that the categorization of variants in **Figure 1** is not specific to this model but instead can be applied much more generally. In principle we can always quantify the per capita growth rate of a variant in a fully naïve and a fully vaccinated population no matter what the epidemiological model or parameter values. The simplification that $\delta_i = 0$ is made entirely for expositional purposes since it allows one to write the overall per capita growth rate of a variant for any level of vaccine coverage as a convex combination of the variant's growth rate in each of the two "pure" populations. As such the growth rates in **Figure 1** are connected simply by straight lines. Furthermore, the kind of categorization in **Figure 1** can be applied regardless of the epidemiological state of the population (e.g., exponential growth, endemicity, etc.). For example, as the epidemic grows the density of uninfected hosts will drop and perhaps non-pharmaceutical interventions will be implemented to reduce the force of infection. Although both processes will reduce the absolute growth rates of all variants we can nevertheless construct a version of **Figure 1** for any epidemiological context of interest. Moreover, because the categorization of variants in **Figure 1** depends only on their *relative* growth rates, if this relative ordering does not change with epidemiological context then a variant's classification into one of the four categories will remain consistent regardless of what is happening epidemiologically.

In the following, for simplicity, the effects of the mutation on the different viral components of fitness in host $X$ (where $X = N$ or $V$) will be assumed to be small and will be denoted:

- $\Delta\beta_X = \beta_m^X - \beta_w^X$
- $\Delta\sigma_X = \sigma_m^X - \sigma_w^X$
- $\Delta\gamma_X = \gamma_m^X - \gamma_w^X$

and the components of fitness of the wildtype will be noted:

- $\delta = \delta_w$
- $\beta_X = \beta_w^X$
- $\sigma_X = \sigma_w^X$
- $\gamma_X = \gamma_w^X$

Using this notation the growth rate of the wildtype population is:

$$r = (1-p)r_N + pr_V$$

and the growth rate of the novel variant is $r + \Delta r$ where:

$$\Delta r = (1-p)\underbrace{(S(\Delta\beta_N\sigma_N + \beta_N\Delta\sigma_N) - \Delta\gamma_N)}_{\Delta r_N} + p\underbrace{(S(\Delta\beta_V\sigma_V + \beta_V\Delta\sigma_V) - \Delta\gamma_V)}_{\Delta r_V}$$

The growth rate of the whole population of all infected individuals is simply:

$$\bar{r} = r + f_m\Delta r \qquad\qquad\qquad \text{(S1)}$$

where $f_m$ is the frequency of the novel variant:

$$f_m = \frac{I_m^N + I_m^V}{I_w^N + I_w^V + I_m^N + I_m^V}$$

Thus, a variant with a higher growth rate will spread when $\Delta r > 0$ and the subsequent increase in mutant frequency will affect the growth rate of the whole pathogen population.

## 3. The dynamics of variant frequency

The dynamics of the variant frequency $f_m$ depends on the distribution of the variant in naïve and vaccinated hosts (Gandon & Day 2007). But if the phenotype of the variant is not very different from that of the wildtype we can obtain a very good approximation of these dynamics using (Otto & Day 2007):

$$\dot{f}_m \approx f_m(1-f_m)\mathbf{V}^T\Delta\mathbf{R}_m\mathbf{F}$$

where $\mathbf{V}^T$ is the vector of reproductive values and $\mathbf{F}$ is the vector of class frequencies which correspond to the conormalised (i.e. $\mathbf{V}^T\mathbf{F} = \mathbf{1}$) left and right eigenvectors of $\mathbf{R}_w$, respectively:

$$\mathbf{F} \propto \left\{ \frac{S_N\sigma_N}{S_V\sigma_V}\left(1 + \frac{\delta}{S_N\beta_N\sigma_N + S_V\beta_V\sigma_V}\right) + O(\delta^2), 1 \right\}$$

$$\mathbf{V}^T \propto \left\{ \frac{\beta_N}{\beta_V}\left(1 + \frac{\delta}{S_N\beta_N\sigma_N + S_V\beta_V\sigma_V}\right) + O(\delta^2), 1 \right\}$$

and selection on the different transitions is given by:

$$\Delta\mathbf{R}_m = \begin{pmatrix} s_{NN} & s_{VN} \\ s_{NV} & s_{VV} \end{pmatrix}$$

where:

$s_{NN} = (\Delta\beta_N\,\sigma_N + \beta_N\,\Delta\sigma_N)\,S_N - \Delta\gamma_N$       Selection coefficient (when $N$ infect $N$)

$s_{NV} = (\Delta\beta_N\,\sigma_V + \beta_N\,\Delta\sigma_V)\,S_V$           Selection coefficient (when $N$ infect $V$)

$s_{VV} = (\Delta\beta_V\,\sigma_V + \beta_V\,\Delta\sigma_V)\,S_V - \Delta\gamma_V$    Selection coefficient (when $V$ infect $V$)

$s_{VN} = (\Delta\beta_V\,\sigma_N + \beta_V\,\Delta\sigma_N)\,S_N$           Selection coefficient (when $V$ infect $N$)

After some calculation this yields:

$$\dot{f}_m \approx f_m(1-f_m)s$$

where:

$$s \propto (1-p)S(\beta_N \Delta\sigma_N + \sigma_N \Delta\beta_N) + pS(\beta_V \Delta\sigma_V + \sigma_V \Delta\beta_V) - (1-q)\Delta\gamma_N - q\Delta\gamma_V + \delta K + O(\delta^2)$$

$$q = \frac{S_V \beta_V \sigma_V}{S_N \beta_N \sigma_N + S_V \beta_V \sigma_V}$$

$$K = \frac{S_N S_V \beta_N \beta_V \sigma_N \sigma_V}{(S_N \beta_N \sigma_N + S_V \beta_V \sigma_V)^2}\left(\left(\frac{\Delta\beta_N}{\beta_N} - \frac{\Delta\beta_V}{\beta_V}\right) + \left(\frac{\Delta\sigma_N}{\sigma_N} - \frac{\Delta\sigma_V}{\sigma_V}\right) - \frac{2}{(S_N \beta_N \sigma_N + S_V \beta_V \sigma_V)}(\Delta\gamma_N - \Delta\gamma_V)\right)$$

When $\delta = 0$ this simplifies as:

$$s \propto \underbrace{(1-p)S(\beta_N \Delta\sigma_N + \sigma_N \Delta\beta_N) - (1-q)\Delta\gamma_N}_{\text{selection in naive hosts}} + \underbrace{pS(\beta_V \Delta\sigma_V + \sigma_V \Delta\beta_V) - q\Delta\gamma_V}_{\text{selection in vaccinated hosts}} \qquad \text{(S2)}$$

Note how selection for higher values of transmission $\beta$ and infectivity $\sigma$ depend on the density of susceptible hosts $S$ while selection on the duration of infection $\gamma$ does not (Day et al 2020).

## 4. The dynamics of differentiation

Next, we use Gandon & Day (2007) to track the difference in variant frequency between vaccinated and naïve hosts. The dynamics of variant frequencies in naïve and vaccinated hosts in a well-mixed population is:

$$\dot{f}_m^N = v_N s_{NN} + v_V \frac{I_V}{I_N} s_{VN} + \frac{I_V}{I_N} \bar{r}_{VN} D$$

$$\dot{f}_m^V = v_V s_{VV} + v_N \frac{I_N}{I_V} s_{NV} - \frac{I_N}{I_V} \bar{r}_{NV} D$$

where:

$D = f_m^V - f_m^N$         Differentiation

$v_N = f_m^N(1 - f_m^N)$      Genetic variance in naive hosts

$v_V = f_m^V(1 - f_m^V)$      Genetic variance in vaccinated hosts

$$\bar{r}_{VN} = f_m^V\big((\beta_V + \Delta\beta_V)(\sigma_V + \Delta\sigma_V)S_N\big) + (1 - f_m^V)(\beta_V \sigma_V S_N)$$
$$= \beta_V \sigma_V S_N + f_m^V(\beta_V \Delta\sigma_N + \Delta\beta_V \sigma_N)S_N = \beta_V \sigma_V S_N + f_m^V s_{VN}$$

$$\bar{r}_{NV} = f_m^N\big((\beta_N + \Delta\beta_N)(\sigma_V + \Delta\sigma_V)S_V\big) + (1 - f_m^N)(\beta_N \sigma_V S_V)$$
$$= \beta_N \sigma_V S_V + f_m^N(\beta_N \Delta\sigma_V + \Delta\beta_N \sigma_V)S_V = \beta_N \sigma_V S_V + f_m^N s_{NV}$$

The dynamics of differentiation $D$ is therefore given by:

$$\dot{D} = v_V\left(s_{VV} - \frac{I_V}{I_N}s_{VN}\right) - v_N\left(s_{NN} - \frac{I_N}{I_V}s_{NV}\right) - D\left(\frac{I_N}{I_V}\bar{r}_{NV} + \frac{I_V}{I_N}\bar{r}_{VN}\right)$$

If we assume there is no differentiation initially ($D = f_m^V - f_m^N = 0$, which also means genetic variance is the same in the two environments, $v_N = v_V = v = f_m(1 - f_m)$) then the dynamics of differentiation are:

$$\dot{D} = v\left(s_{VV} - s_{NN} - \frac{I_V}{I_N}s_{VN} + \frac{I_N}{I_V}s_{NV}\right)$$

$$\dot{D} = v\bigg( (\Delta\beta_V\ \sigma_V + \beta_V\ \Delta\sigma_V)\,S_V - \Delta\gamma_V - (\Delta\beta_N\ \sigma_N + \beta_N\ \Delta\sigma_N)\,S_N + \Delta\gamma_N$$

$$- \frac{I_V}{I_N}(\Delta\beta_V\ \sigma_N + \beta_V\ \Delta\sigma_N)\,S_N + \frac{I_N}{I_V}(\Delta\beta_N\ \sigma_V + \beta_N\ \Delta\sigma_V)\,S_V \bigg)$$

If we further assume that the prevalence is low so that $S_N$ and $S_V$ remains constant during the early stage of the epidemic the prevalence will grow exponentially and the ratio $\frac{I_N}{I_V}$ will remain constant. The value of this ratio can be computed from the vector $\mathbf{F}$ of class frequencies given above:

$$\frac{I_N}{I_V} = \frac{S_N\sigma_N}{S_V\sigma_V}\left(1 + \frac{\delta}{S_N\beta_N\sigma_N + S_V\beta_V\sigma_V}\right) + O(\delta^2)$$

The dynamics of differentiation therefore becomes:

$$\dot{D} = v\bigg( (S_N\beta_N\sigma_N + S_V\beta_V\sigma_V)\left(\frac{\Delta\sigma_V}{\sigma_V} - \frac{\Delta\sigma_N}{\sigma_N}\right) + \Delta\gamma_N - \Delta\gamma_V + \frac{s_{NV}\frac{\sigma_V S_V}{\sigma_N S_N} + s_{VN}\frac{\sigma_N S_N}{\sigma_V S_V}}{S_N\beta_N\sigma_N + S_V\beta_V\sigma_V}\delta \bigg)$$

$$+ O(\delta^2)$$

When $\delta = 0$ this simplifies as:

$$\dot{D} = v\left(S\big((1-p)\beta_N\sigma_N + p\beta_V\sigma_V\big)\left(\frac{\Delta\sigma_V}{\sigma_V} - \frac{\Delta\sigma_N}{\sigma_N}\right) - (\Delta\gamma_V - \Delta\gamma_N)\right) \qquad \text{(S3)}$$

Note how differentiation is not driven by the transmission rates of the mutant but by its relative infectivity in naïve and vaccinated hosts.

**Figure Caption**

**Figure S1: Typology of pathogen variants after vaccination.** We can identify 8 different types of variants. The panel (**a**) is expanding the description of **Figure 1** and the panel (**b**) is indicating the location of these 8 types as in **Figure 2**. Variant type I is adapted to naïve hosts but maladapted on vaccinated hosts. Variant type V is maladapted on both types of hosts. We focus on the 6 vaccine-adapted variants with $\Delta r_V > 0$. Variants II, III and IV are generalist variants (i.e., $\Delta r_N > 0$) and the magnitude of $\Delta r_V$ explains the difference between these 3 variants. Variants VI, VII and VIII are specialist variants (i.e., $\Delta r_N < 0$) and the magnitude explains the difference between these 3 variants. Note that variants IV,VII and VIII have a growth rate that increases with vaccination coverage. This increased growth rate can have major public health implications. In particular, with variants IV and VIII, evolution is expected to yield a higher pathogen growth rate after 100% vaccination (the evolved growth rate $r_V$ is indicated with the black dot) than after 0% vaccination (the evolved growth rate $r_N$ indicated with the white dot).

# Appendix 2

In Figure 2 we use the per capita growth rate, $r$, of infections in a fully naïve and a fully vaccinated population as the axes for visualizing different variants. Another common measure of the fitness of a variant is its reproduction number, $R_0$. The reproduction number $R_0$ is a dimensionless quantity that gives the number of new infections produced by a single infected individual. It does not account for the timing of when these infections occur and so it does not provide information about the timeframe over which infections are spreading through a population. The per capita growth rate has units 1/time and so it does account for the timeframe of infection spread. Two variants can therefore have different values of $r$ even if they each produce the same total number of $R_0$ new infections per infection, if they differ in how the production of these new infections is spread out over time. More specifically, if we use $t$ to denote the amount of time that has elapsed since the start of an infection, and $\phi(t)dt$ to denote the fraction of the $R_0$ new infections that are produced in the time interval $(t, t + dt)$, then $r$ and $R_0$ are related by the equation $R_0 = 1/\int_0^\infty e^{-rt}\phi(t)dt$ (Wallinga and Lipsitch 2007). In the special case where $\phi(t)$ is a gamma density with mean $T$ and parameter $k$, we can solve this for $r$ to obtain

$$r = k\left(R_0^{1/k} - 1\right)/T \tag{S4}$$

For the model of **Appendix 1** we have $R_{0,i}^X = \frac{s\beta_i^X\sigma_i^X}{\delta+\gamma_i^X}$ as the reproduction number of variant $i$ in host type $X$ (where $X = N$ or $V$), $T_i^X = \frac{1}{\delta+\gamma_i^X}$ as the mean time during an infection when new infections are produced, and $k = 1$.

There are several alternative ways one might construct plots analogous to that of **Figure 2**. One is simply to use a variant's reproduction number in each of the two host types as the measure of fitness rather than its per capita growth rate. However, doing so has two disadvantages. First, differences between variants in the timing of transmission will affect their relative competitive abilities during a vaccination campaign and this is not accounted for when using the reproduction number as a measure of fitness (for the reasons discussed above). Second, as has become clear with the appearance and spread of novel variants during the SARS-CoV-2 pandemic, one of the first and easiest quantities estimated for a variant is its selection coefficient (i.e., the difference in growth rate between it and the current dominant type: $s = r_{new} - r_{old}$). Such estimates are all that is required to place a new variant on a plot like **Figure 2** (for instance see below how we can attempt to place Omicron in Figure 5 using its selection coefficient against Delta).

Another alternative to the axes used in **Figure 2** would be to use a variant's overall fitness across both host types as one axis (as either $r$ or $R_0$) and its ability to evade vaccine-induced immunity as the other. The latter is often quantified by the vaccine efficacy against the variant, as measured by $VE_i = 1 - \frac{\sigma_i^V}{\sigma_i^N}$, where $\sigma_i^N$ and $\sigma_i^V$ is the infectivity of the variant in naïve and vaccinated hosts respectively. There are drawbacks with this approach as well. First, the axes in such a plot are not independent. Second, any measure of the overall fitness of a variant will necessarily depend on vaccination coverage and so a different plot would be needed for different vaccination coverages. It is possible to circumvent this problem by using the fitness of a variant in the naïve population as one of the axes rather then its overall fitness, but then it is no longer possible to visualize overall fitness on such a plot. The reason is that a combination of a variant's fitness in naïve hosts and its $VE_i$ is not sufficient to determine its overall fitness. Differences between variants in how the vaccine affects disease severity and transmissibility can also influence a variant's competitive ability and these are excluded from such a plot (since $VE_i$ quantifies only the effect on infectivity which is only one of three main components of pathogen fitness, see **Box 1**).

In the remainder of this Appendix we use currently available estimates from three distinct variants to obtain the values of the per capita growth rates in naïve and vaccinated hosts plotted in **Figure 5**. For simplicity, we assume that the generation time is the same for all variants in naïve and vaccinated hosts (Ganyani et al. 2020). The following table summarises the estimates used to make **Figure 5**. We have attempted to account for the uncertainty in the estimates (within and between studies) by indicating a range of parameter estimates in the following table (min and max). We also note that although estimates of vaccine efficacy against different variants vary among studies, their relative ranking is typically consistent ($VE_w > VE_\alpha > VE_\delta$) (see CDC Science Brief). Finally, the estimates used to construct **Figure 5** are based on the above formula that relates the reproduction number of the per capita growth rate, and use a value $k$ = 5.

**Table S1:** Estimates of the life-history characteristics of the different variants. We present the estimates used for the plots in **Figure 5** as well as a range of values that were reported among different studies. Numbers in parenthesis indicate the minimal and maximal values used to plot the horizontal and vertical bars in **Figure 5**.

|  |  | $R_{0,i}^N$ | $T_i^N = T_i^V$ (days) | $\dfrac{\beta_i^V}{\beta_i^N}$ | $\dfrac{\sigma_i^V}{\sigma_i^N}$ | $r_i^N$ (days$^{-1}$) | $r_i^V$ (days$^{-1}$) |
|---|---|---|---|---|---|---|---|
| Variants | Wuhan | **2.5** (2,3) | **5** | **0.5** (0.4,0.6) | **0.1** (0.05,0.2) | **0.20** (0.15,0.25) | **-0.34** (-0.47,-0.18) |
|  | Alpha | **4.25** (3.5,5.5) | **5** | **0.5** (0.4,0.6) | **0.15** (0.05,0.25) | **0.34** (0.28,0.38) | **-0.20** (-0.41, -0.06) |
|  | Delta | **7** (5,9) | **5** | **0.5** (0.4,0.6) | **0.2** (0.1,0.4) | **0.48** (0.38,0.55) | **-0.07** (-0.28, 0.17) |

At the time of writing a new SARS-CoV-2 variant of concern labelled Omicron has begun a selective sweep in several parts of the world. At present there are multiple estimates of its selection coefficient in different geographic regions, and these are between 0.2 and 0.4 per day. There is also growing evidence that current vaccines provide much less protection against Omicron than they do against Delta, but precise estimates are still lacking. As such, at the moment we can use the estimate of the selection coefficient to tentatively place Omicron somewhere in the ellipse shown in **Figure 5**. This ellipse was constructed so that the main axis is aligned with a contours of equal fitness (i.e., a line with a slope equal to $-p/(1-p)$ with $p = 0.7$, see **Figure 3**) and a selection coefficient $s = r_{Omicron} - r_{Delta}$ of 0.3.

A

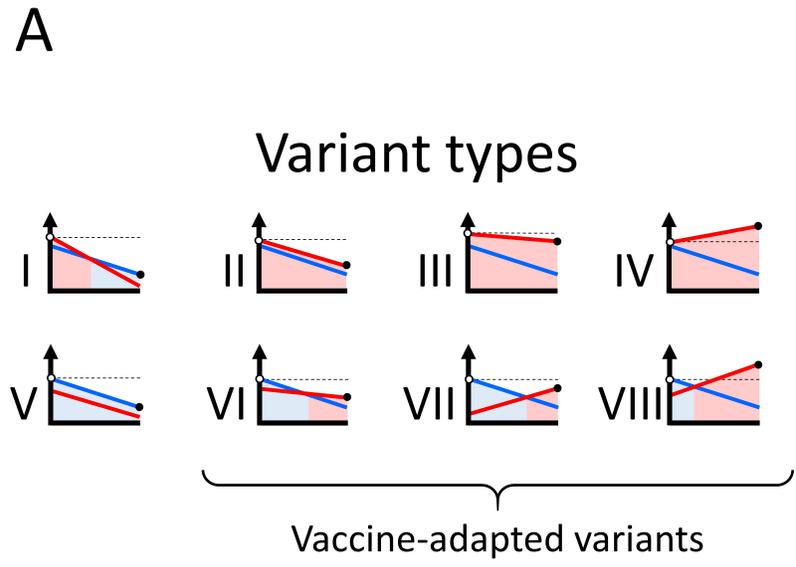

## Variant types

I

II III IV

V

VI VII VIII

Vaccine-adapted variants

B

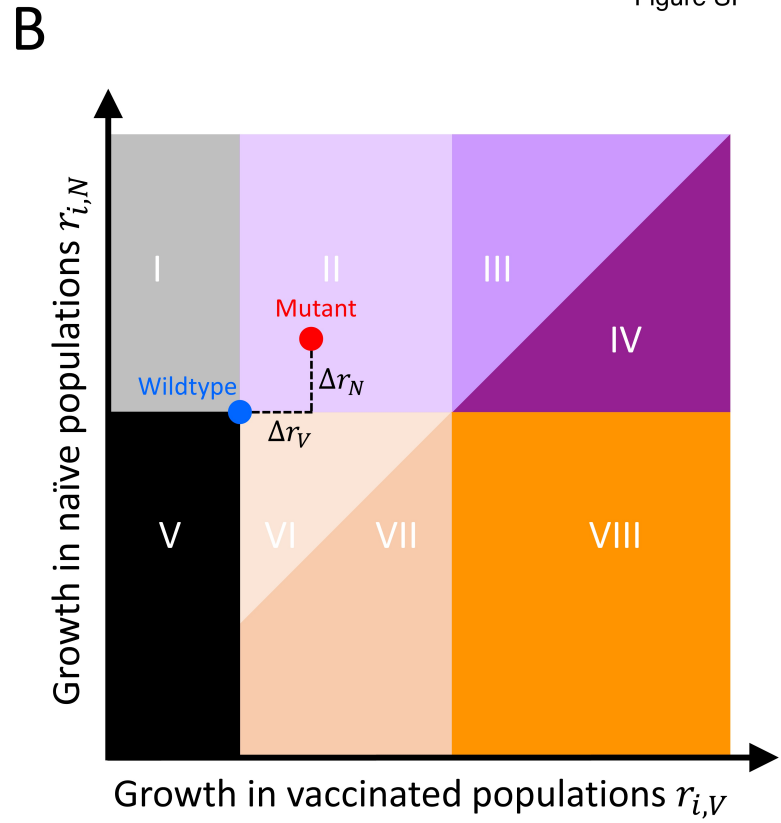

Growth in vaccinated populations $r_{i,V}$